# Excitations in the field-induced quantum spin liquid state of α-RuCl₃

**Authors:**

A. Banerjee*[1], P. Lampen-Kelley*[2,3], J. Knolle[4], C. Balz[1], A.A. Aczel[1], B. Winn[1], Y. Liu[1], D. Pajerowski[1], J.-Q. Yan[2], C.A. Bridges[5], A.T. Savici[6], B. C. Chakoumakos[1], M. D. Lumsden[1], D.A. Tennant[7], R. Moessner[8], D.G. Mandrus[2,3], S.E. Nagler[1]

**Affiliations:**

[1]Quantum Condensed Matter Division, Oak Ridge National Laboratory, Oak Ridge, TN 37831, U.S.A.

[2]Materials Science and Technology Division, Oak Ridge National Laboratory, Oak Ridge, TN, 37831, U.S.A.

[3]Department of Materials Science and Engineering, University of Tennessee, Knoxville, TN 37996, U.S.A.

[4]Department of Physics, Cavendish Laboratory, JJ Thomson Avenue, Cambridge CB3 0HE, U.K.

[5]Chemical Sciences Division, Oak Ridge National Laboratory, Oak Ridge, TN, 37831, U.S.A.

[6]Neutron Data Analysis and Visualization Division, Oak Ridge National Laboratory, Oak Ridge, TN 37831, U.S.A.

[7]Neutron Sciences Directorate, Oak Ridge National Laboratory, Oak Ridge, TN 37831, U.S.A.

[8]Max Planck Institute for the Physics of Complex Systems, D-01187 Dresden, Germany.

The Kitaev model on a honeycomb lattice predicts a paradigmatic quantum spin liquid (QSL) exhibiting Majorana Fermion excitations. The insight that Kitaev physics might be realized in practice has stimulated investigations of candidate materials, recently including $\alpha$-RuCl$_3$. In all the systems studied to date, significant non-Kitaev interactions induce magnetic order at low temperature. However, in-plane magnetic fields of roughly 8 Tesla suppress the long-range magnetic order in $\alpha$-RuCl$_3$ raising the intriguing possibility of a field-induced QSL exhibiting non-Abelian quasiparticle excitations. Here we present inelastic neutron scattering in $\alpha$-RuCl$_3$ in an applied magnetic field. At a field of 8 Tesla the spin waves characteristic of the ordered state vanish throughout the Brillouin zone. The remaining single dominant feature of the response is a broad continuum centered at the $\Gamma$ point, previously identified as a signature of fractionalized excitations. This provides compelling evidence that a field-induced QSL state has been achieved.

The Kitaev model on a honeycomb lattice [1] has been exactly solved to reveal a unique quantum spin liquid (QSL) exhibiting itinerant Majorana Fermion and gauge-flux excitations. The Kitaev candidate system α-RuCl$_3$ is an insulating magnetic material comprised of van der Waals coupled honeycomb layers of $4d^5$ Ru$^{3+}$ cations nearly centered in edge-sharing RuCl$_6$ octahedra. A strong cubic crystal field combined with spin-orbit coupling leads to a Kramer's doublet, nearly perfect J = 1/2 ground state [2-4], thus satisfying the conditions necessary for producing Kitaev couplings in the low energy Hamiltonian [5]. Similar to the widely studied honeycomb [6] and hyper-honeycomb [7] Iridates, at low temperatures α-RuCl$_3$ exhibits small-moment antiferromagnetic zigzag order [3, 8-11] with T$_N$ ≈ 7 K for crystals with minimal stacking faults. In the zigzag state the magnetic excitation spectrum shows well-defined low-energy spin waves with minima at the M points (See Supplementary Materials (SM) Fig. S1 for the Brillouin Zone (BZ) definition) as well as a broad continuum that extends to much higher energies centered at the Γ points [12, 13]. Above T$_N$ the spin waves disappear but the continuum remains, essentially unchanged until high temperatures of the order of 100 K [3, 12, 13]. In analogy with the situation for coupled spin- ½ antiferromagnetic Heisenberg chains [14], the high energy part of the continuum has been interpreted as a signature of fractionalized excitations [3, 12, 13]. The overall features of the inelastic neutron scattering (INS) response resemble those of the Kitaev QSL [15-17] and are consistent with an unusual response seen in Raman scattering [16, 18, 19], suggesting that the system is proximate to a QSL state exhibiting magnetic Majorana fermion excitations [3, 12, 13]. Magnetic field offers a clean quantum tuning parameter for Kitaev materials [7-9, 20] and can be applied on large single crystals facilitating INS studies. It is known to suppress the magnetic order in α-RuCl$_3$ [8, 9, 21-26] raising the intriguing possibility of a field-induced QSL [20]. It is thus of great interest to investigate the nature of the excitations in the magnetic field-induced disordered state.

To investigate this phenomenon, high-quality single crystals of α-RuCl$_3$ were grown using vapor-transport techniques [3, 10]. Figure 1 shows bulk susceptibility and neutron diffraction measurements, demonstrating the suppression of the zigzag order (Fig. 1a) when the field is applied along the ζ = (-1, 2, 0) (trigonal notation, see e.g., ref. 10-12) or equivalent directions (Fig. 1b). The cusp in the susceptibility occurs at T$_N$ = 7.5 K at 0.1 T, close to the location of the zero-field heat capacity anomaly in single-phase RuCl$_3$ crystals with ABC magnetic stacking [10]. With increasing field, T$_N$ shifts to lower temperatures and the transition is not observed beyond B$_C$ = 7.3 (3) T.

Magnetic Bragg peaks associated with the zigzag spin order appear below T$_N$ at (½, 0, L), (0, ½, L) and (½, -½, L) where L ≠3$n$ [9-12, 21]. Consistent with Sears *et al.* [21], magnetic fields of 2 T along {1 1 0}-equivalent directions completely suppress the intensity of magnetic Bragg peaks with **Q** ⊥ **B** as seen in Fig. 1c. Conversely, peaks with a significant projection of **Q** along **B** gain intensity at low fields with increasing fields. This intensity redistribution can signify a reorientation of the ordered moments to lie perpendicular to the field direction, or a depopulation of domains with moments aligned along the field direction [21]. The weighted average intensity of magnetic peaks with L = 1 is roughly flat up to B ≈ 3.5 T and then follows a downward trend. (See Figs. S2 and S3 for further details on the behavior of (½, 0, 1) and (½, 0, 2) orders with applied field to 2 T).

Figure 2 shows time-of-flight INS from a 740 mg single crystal mounted with the (H, 0, L) scattering plane horizontal, and a vertical magnetic field applied in the $\zeta$ = (-1, 2, 0) direction (see Methods). The scattering at 2 K in zero field (Fig. 2a) shows low-energy, gapped spin waves with the expected minima at the (±1/2, 0, L) M points. The spin waves also show local minima at the $\Gamma$ (H = 0) and Y (H = 1) points. The scattering near the $\Gamma$ point, however, is dominated by a broad continuum extending to higher energies, consistent with previous measurements [3, 12, 13].

Figures 2b-e, show the evolution of the spectrum at 2 K in 2 T increments from B = 2 to 8 T. At B = 2 T the spin wave intensity is increased. This is consistent with expectations for the neutron scattering cross-section. As noted above, one likely explanation (Fig. 1c) is at 2 T the ordered moment direction becomes perpendicular to the applied field, and nearly parallel to (1/2, 0, 1). This naturally leads to a decreased intensity for (1/2, 0, 1) peaks and an enhanced scattering intensity for the concomitant M-point spin waves (Fig. 2b) since the later represent fluctuations of components perpendicular to the ordered moment. The spin waves persist as the field is increased to 4 T (Fig. 2c), and begin to lose intensity by 6 T (Fig. 2d). At 8 T, above $B_C$, spin-wave scattering is totally suppressed and the intensity of the continuum is enhanced (Fig. 2e). The response is an intense, and apparently gapped, column of scattering at the $\Gamma$ point. Over most of the energy range, the 8 T spectrum bears strong resemblance to the zero-field spectrum just above $T_N$ (Fig. 2f) [12, 13].

Further details of the in-field spectra are presented in Fig. 3. Constant-Q cuts at the M point between 0 and 6 T in Fig. 3a show a spin wave at E = 2.25 ± 0.11 meV, consistent with previous zero-field experimental results [12, 27]. The cuts verify that the spin-wave intensity is enhanced at 2 T, gradually reduced by 6 T, and completely absent at 8 T, consistent with the suppression of magnetic order above $B_C$.

Figure 3b shows constant-Q cuts at the zone center ($\Gamma$ point, H = 0). The zero-field cut at 2 K exhibits a peak at E = 2.69 ± 0.11 meV, consistent with a recent observation of the zone-center magnon by THz spectroscopy [28]. The scattering is largely unchanged up to 4 T. Conversely, at 6 T the low energy scattering has "filled-in", suggesting that the energy gap at the $\Gamma$ point has closed. On further increasing the field to 8 T, the scattering consists of a broad peak centered around 3.4 meV that merges into the higher energy continuum. Extrapolation of the low-energy $\Gamma$-point spectrum suggests a re-opening of a new gap by 8 T. The appearance of a field-induced gap in the spectrum of spin excitations above $B_C$ has been observed in NMR [26] and inferred from thermal transport [24] measurements.

Figure 3c plots the wave-vector dependence of the T = 2 K scattering at different fields, integrated over the energy interval [5, 7] meV. As seen previously [12, 13], the scattering profile is a peak centered at the $\Gamma$ point. At low temperatures and low fields the peak width is broader, reflecting contributions associated with enhanced correlations related to the zigzag order [12]. A similar comparison of the T = 2 K, B = 8 T with the T = 15 K, zero-field scattering in Fig. 3d demonstrates near quantitative agreement between the two magnetically disordered states. This is true over the available data range for energy transfers above the maximum of the spin wave band.

It has been shown previously [12, 13] that the energy continuum observed at the $\Gamma$ point in zero field both above and below $T_N$ is similar to the T = 0 response function of a Kitaev QSL [15-17], and therefore may be a signature of fractionalized excitations associated with proximity to the QSL. Combined exact diagonalization and DMRG calculations [20] for extended Kitaev-Heisenberg Hamiltonians have provided indications for a magnetic field-induced transition from zigzag order to a gapped QSL state. The disappearance of spin waves at $B_C$, combined with the appearance of a gap in the continuum excitation spectrum that is unconnected with spin waves provides significant evidence for such a field-induced QSL in $\alpha$-RuCl$_3$. This observation is consistent with the interpretation of NMR experiments by Baek *et al.* [26].

The removal of magnetic long-range order by the field enables the comparison of the measured energy-dependent scattering to a QSL-based theory over a large portion of the bandwidth of the magnetic excitations. The full effective magnetic Hamiltonian describing $\alpha$-RuCl$_3$ has not been definitively determined so that the concomitant dynamic response functions remain unknown. In view of this, as a starting point it is reasonable to compare the results to exact calculations for a Kitaev QSL. Previous such comparisons [3, 12] of the zero-field measurements were restricted to high-energy features with relatively small spectral weight. The expected scattering is broad in energy for both ferromagnetic (FM) and antiferromagnetic (AFM) Kitaev QSLs, however in zero field the momentum dependence of the scattering was seen to be similar to the response calculated for an AFM

Kitaev QSL at T = 0. With a phenomenological extension of the calculations to nonzero T, the full energy dependence of the intensity at the Γ point appears to be closer to the response calculated for a FM Kitaev model in an effective magnetic field (see supplementary materials, Figs. S5, S6). The latter exhibits the following features that are qualitatively consistent with experiment: For temperatures comparable to, or larger than, the flux gap, the signal near **Q** = 0 is a broad continuum, exhibiting only a moderately intense peak at the lower-energy threshold. Secondly, the high frequency part of the spectrum is resilient as a function of temperature (up to $T \sim J_K$, the Kitaev constant) or magnetic field (both below and above $B_c$). Finally, with field, the low frequency response acquires a low-energy gap with an intensity enhanced at higher energies, similar to the Γ point continuum scattering seen in α-RuCl$_3$ above $B_c$ (Fig. 3b). This fits with the idea that the field-induced QSL evolves from the zero-field Kitaev QSL as time-reversal symmetry breaking opens a gap in the Majorana spectrum and a Majorana flux bound state (broadened by the presence of thermally excited fluxes at nonzero temperature) enhances the low frequency response.

The overall similarity between the excitation spectra for $T > T_N$ (B = 0 T) and $B > B_C$ (T = 2 K) is quite remarkable (see Fig. 2e, f and Fig. 3d), and suggests the possibility of a simple connection between the high-energy excited states in the two regions of parameter space. Generally (see e.g., ref. 29) one would expect Zeeman splitting to drive a softening of the gapped spin wave mode at the zigzag ordering wavevector (M point) as the applied field is increased, eventually driving the system to a phase transition when the gap is closed at the ordering wave vector. Here the lack of any observed splitting or an M point gap softening, coupled with the apparent softening of the gap at the Γ point, is surprising, and suggests the possibility of another field-induced transition between the zigzag state and the QSL seen above $B_C$. Figure 3e shows a plot of the intensity and FWHM of the scattering continuum as a function of field. The results suggest an anomaly or discontinuity within the shaded region, in the vicinity of 6 T. Further evidence is provided by the isothermal (T = 2 K) AC susceptibility shown in Fig. 3f. This shows a large anomaly at $B_C$, with a second anomaly near 6 T. Whether or not this indicates a second transition, and thus the presence of an intermediate phase between the ordered magnet and a field-induced QSL is the subject of further investigation.

To conclude, we have shown that with a magnetic field applied in the {110} direction, long-range magnetic order in $\alpha$–RuCl$_3$ disappears above a threshold field $B_C$, and the high-frequency spectrum of magnetic excitations for B > $B_C$ resembles that expected for a Kitaev QSL. This QSL is expected to be topologically different from the zero-field Kitaev QSL, and hence the next challenge is to identify and explore experimentally its most compelling signatures. These may include topologically protected edge states and quasiparticle excitations with non-Abelian statistics [1], which have generated much enthusiasm about topological quantum computation [30-32].

**Methods:**

**Synthesis and bulk characterization:** Single crystals of α-RuCl$_3$ were prepared using vapor-transport techniques from pure α-RuCl$_3$ powder as described previously [10, 12]. Crystals grown by the same method have been extensively characterized via bulk and neutron scattering techniques [10, 12]. All samples measured in the current work exhibit a single magnetic phase at low temperature with a transition temperature Tc ∼ 7 K, indicating high crystal quality with minimal stacking faults [10]. DC magnetization and AC susceptibility in 12 mg and 17 mg single crystals, respectively, were collected in DC fields of up to 14 T in a Quantum Design Physical Property Measurement System (PPMS). The magnetic field was applied along the reciprocal {1 1 0} directions as identified by Laue diffraction as described in Fig. 1b.

**Neutron diffraction experiments:** Elastic neutron studies in a 5 T vertical-field cryomagnet were performed at Spallation Neutron Source (SNS), Oak Ridge National Laboratory (ORNL), using the CORELLI beamline [33]. CORELLI is a time-of-fight instrument where a pseudo-statistical chopper separates the elastic contribution. A 125 mg α-RuCl$_3$ crystal was mounted on an Al plate and aligned with the (H, 0, L) plane horizontal and B along the ζ = (-1, 2, 0) vertical direction (See Fig. 1b). The crystal was rotated through 170 degrees in 2° steps. Perpendicular coverage of ± 8° (limited by the magnet vertical opening) allowed access to the set of magnetic Bragg peaks of the zigzag ordered phase at the M-points within the first BZ - at (±½, 0, L) in the (H, 0, L) plane and (±½,∓½, L) and (0, ±½, L) out of the (H, 0, L) plane with L=±1, as well as the out-of-plane peaks with L= ±2. The data were reduced using Mantid [34]. Diffraction measurements were also obtained on the same 740 mg crystal measured used in the inelastic study using the HB1A and HB3 triple axis instruments at the High Flux Isotope

Reactor (HFIR). For both HFIR measurements the sample was aligned with the (H, 0, L) scattering plane horizontal and with an applied vertical field of up to 5 T. An incident energy $E_i$=14.7 meV was used. Detailed mesh scans and order parameter measurements confirmed that the magnetic phase transition remains sharp as the field is increased from zero (Fig. S3) and the in-plane magnetic peaks remain at the commensurate position.

**Inelastic scattering experiments:** Single-crystal inelastic neutron scattering measurements were carried out on a 740 mg $\alpha$-RuCl$_3$ crystal in an 8 T vertical-field cryomagnet using the HYSPEC instrument at SNS [35]. The sample was aligned in the horizontal (H, 0, L) scattering plane, with the magnetic field (**B**) parallel to the vertical $\zeta = (-1, 2, 0)$ direction (See Fig. 1b). An incident energy of $E_i$ = 17 meV combined with a Fermi chopper frequency of 240 Hz yielded an experimental energy resolution FWHM = 0.88 ± 0.03 meV based on a Gaussian fit on the elastic line. The zero-field data at 15 K was obtained using a Fermi chopper frequency of 120 Hz, which increases the intensity, but decreases the energy resolution, both by a factor of 2, as compared to 240 Hz. The 15 K data in Fig. 3(d) is scaled by the factor of 2 to allow a direct comparison of the intensity with the 2 K data taken at 240 Hz. These settings provide reasonable (Q, E) coverage and resolution with which to examine the spin-wave spectrum in the (H, 0, 0) direction, however using $E_i$ = 17 meV, the higher-energy part of the sample spectrum (i.e., above roughly 6 meV) is limited by kinematic restrictions. Data were collected in 1° steps as the sample was rotated through 200° about the vertical axis for every temperature and field condition. Empty aluminum sample holder measurements were performed under identical conditions as the sample measurements and the resultant scattering has been subtracted from all data presented in the manuscript. Reduction of the raw data was carried out using standard data reduction routines and Python codes available within Mantid software [34].

**Figures:**

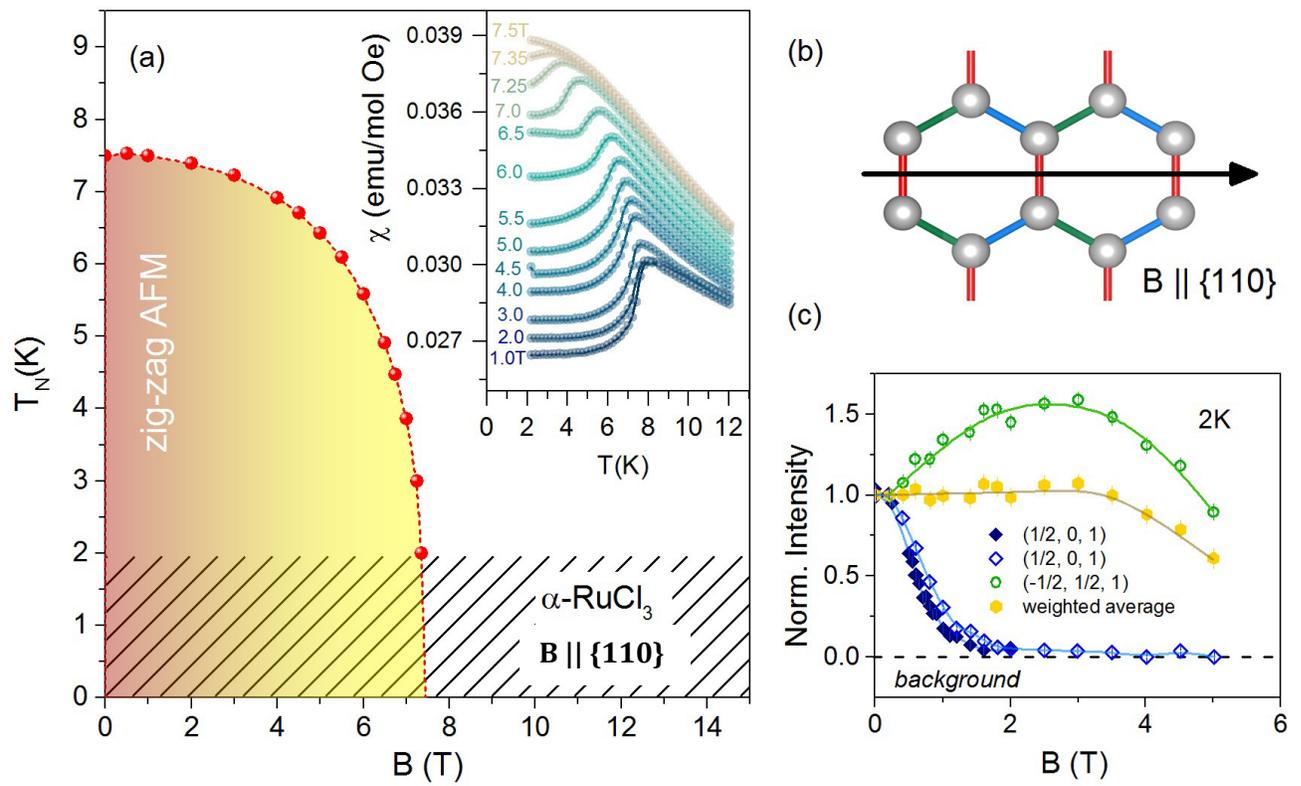

**Figure 1: Field-evolution of the zigzag order in α-RuCl₃:** Bulk susceptibility and diffraction are shown as a function of magnetic field (B) applied along a reciprocal {110} direction. Note that the {110} is equivalently one of (1, 1, 0), (-1, 2, 0) or (2, -1, 0) directions. **(a)** Néel temperature vs. B determined from the cusp in temperature-dependent magnetic susceptibility curves measured in fixed fields **(inset)**. For each field, χ (T) was collected on warming after cooling in zero magnetic field. The region below the experimental base temperature is indicated by hash marks. **(b)** Schematic of the magnetic field direction in the honeycomb plane. **(c)** Magnetic Bragg peak intensity normalized to the zero field value (i.e., I(B)/I(B=0)). The field is applied in the (-1, 2, 0) direction perpendicular to the (H, 0, L) plane. Blue diamonds show the suppression of the in-plane (1/2, 0, 1) magnetic peaks in two different crystals measured at HYSPEC (solid symbols) and CORELLI (open symbols). The green circles show the out-of-plane magnetic peak at (-1/2, 1/2, 1) measured at CORELLI (See Methods for instrument parameters). The yellow symbols represent the weighted average intensity over magnetic peaks with L = 1. In this average, the peaks within the (H, 0, L) plane contribute 1/3 of the intensity. The solid lines are visual guides.

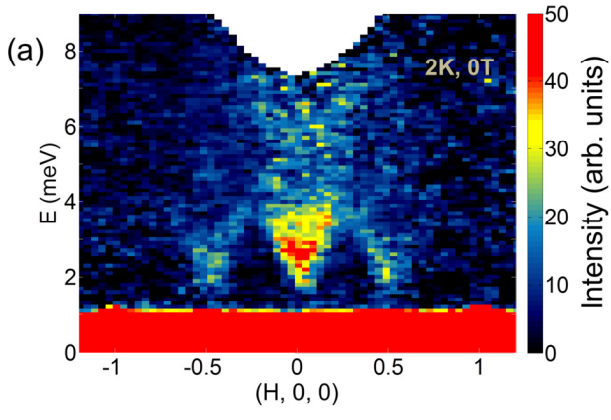
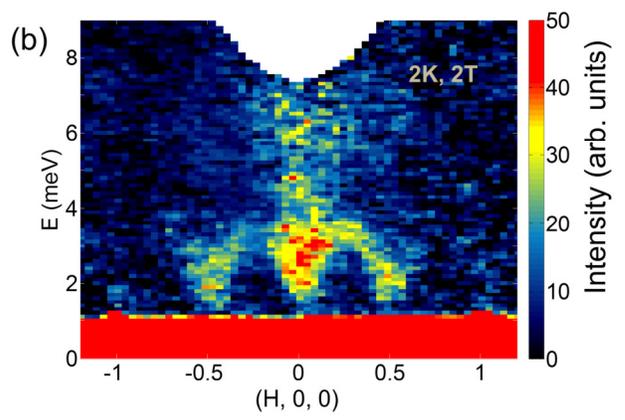
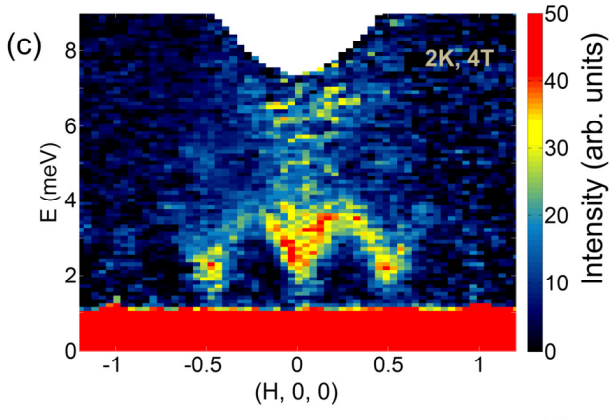
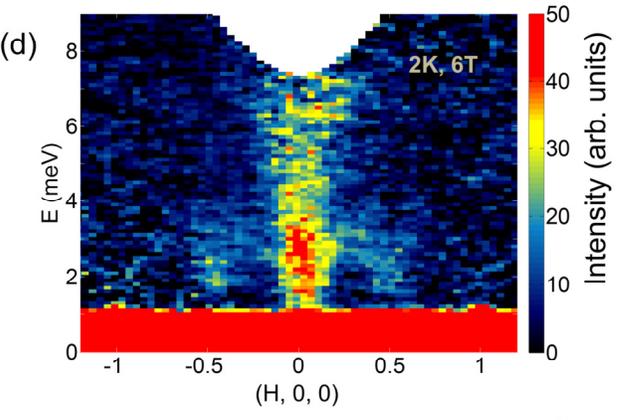
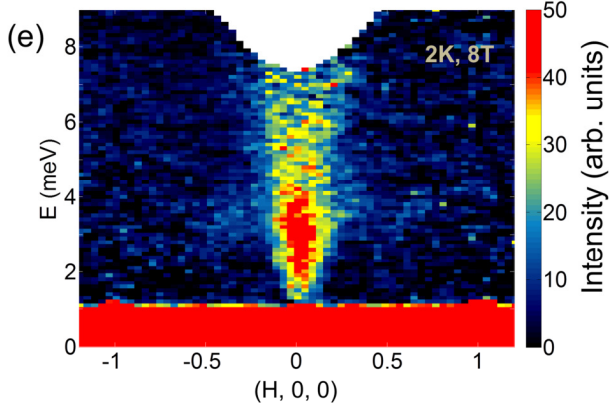
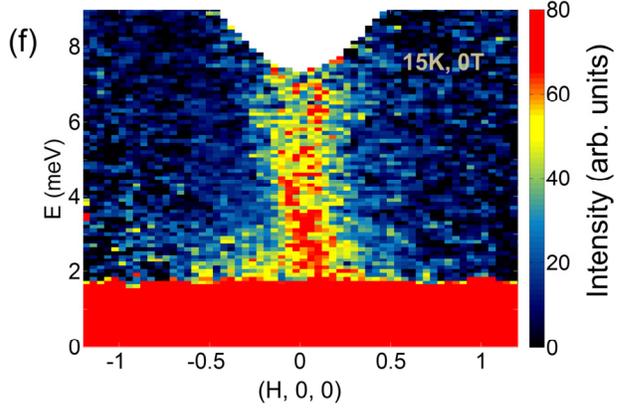

**Figure 2: Field and temperature dependence of INS along (H, 0, 0).** Data obtained on a 740 mg single crystal α-RuCl$_3$ using the HYSPEC chopper spectrometer with Ei = 17 meV (see Methods). **(a)-(e)** Measurements at T = 2 K with an external field applied along the ζ=(-1, 2, 0) direction (perpendicular to (H, 0, 0)) using a Fermi chopper frequency of 240 Hz. The field strengths are **(a)** 0 T, **(b)** 2 T, **(c)** 4 T, **(d)** 6 T, and **(e)** 8 T. The slices shown are integrated over ranges Δζ=[-0.05, 0.05] (in units of 2.1 Å$^{-1}$, see Fig. S1) and ΔL = [-2,2] (in units of 0.37 Å$^{-1}$). The energy continuum at the Γ point is visible at all fields. The low-field data clearly shows the spin-wave spectrum along the (H, 0, 0) direction, with minima at the special points, Y-M-Γ-M-Y, of the 2D honeycomb BZ (see Fig S1 for BZ definitions). **(f)** The zero-field spectrum at T = 15 K, using a Fermi chopper frequency of 120 Hz (see Methods).

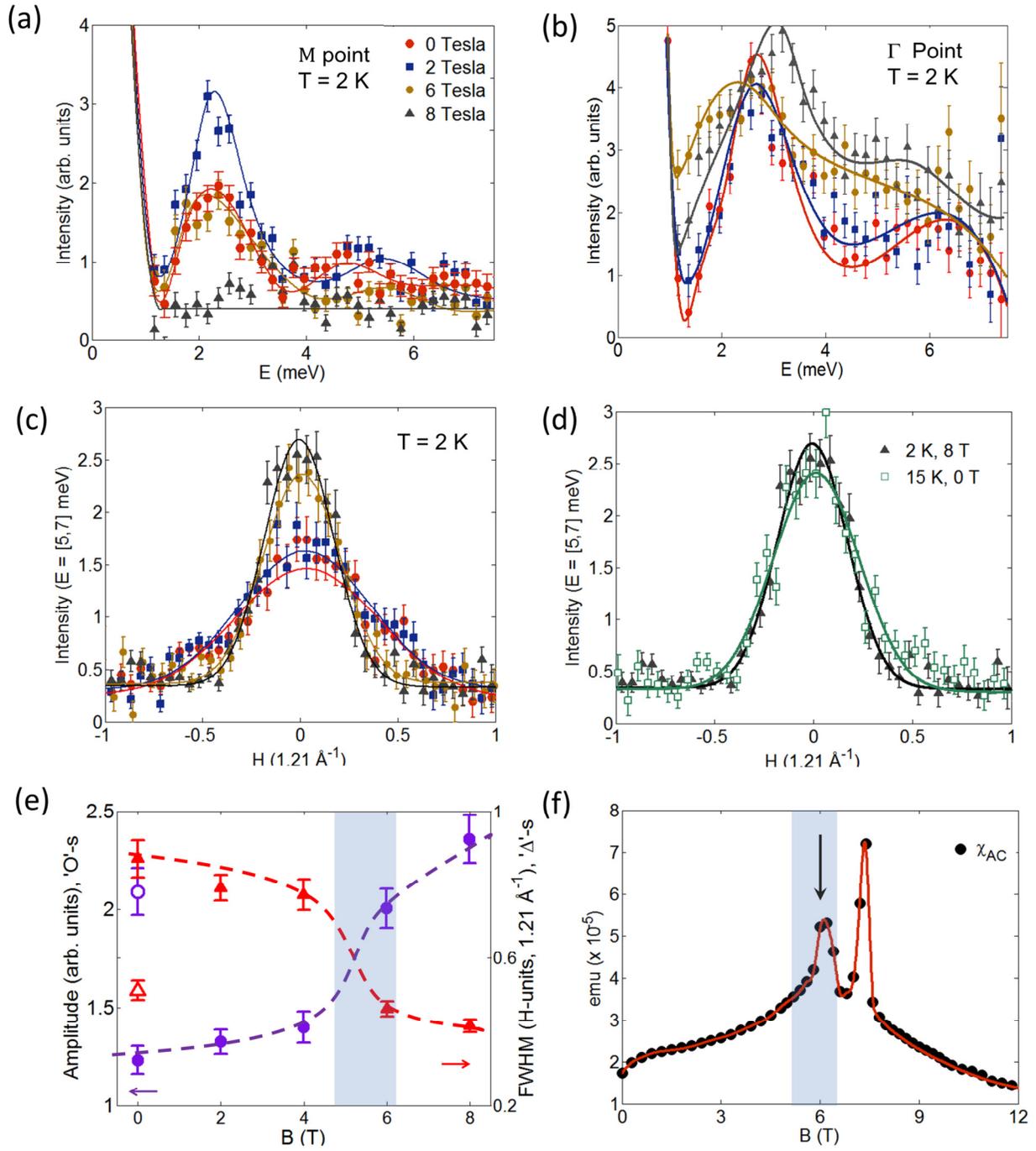

**Figure 3: Detailed field dependence of the response:** The field is applied in the direction of ζ = (−1, 2, 0), and field values are defined in the legend of panel (a). **(a) & (b)** Constant Q cuts showing the energy dependence of the scattering. The data are integrated over the wave vector ranges ΔH = [-0.1, 0.1], Δζ = [-0.05, 0.05] and ΔL = [-2, 2] (See Fig. S1 for units). Lines are guides to the eye. **(a)** Scattering at the M point (1/2, 0, L) shows that the energy of the spin-wave peak at 2.2 ± 0.2 meV does not change with fields up to 6 T. At 8 T the spin waves are suppressed. Defining the energy gap as the point where inelastic scattering exceeds the background leads to value of at least 1.6 ± 0.2 meV. The gap at the M point is also independent of fields up to 6T. **(b)** Scattering at the Γ point showing the evolution of the spectrum. The lowest energy zero-field spin-wave peak occurs at 2.69 ± 0.11 meV. The energy gap for the 0 – 4 T spectra is 1.8 ± 0.2 meV. The gap appears closed at 6 T, but has reopened at 8 T. At 8 T the spectrum shows higher intensity for all energies above the gap. **(c)** T = 2 K constant energy cuts integrated over Δζ and ΔL ranges as described above, and energy range E = [5, 7] meV, showing the evolution of the peak height and width of the continuum at 0, 2, 6 and 8 T. **(d)** The same cuts comparing the data at T = 2 K, 8 T (grey triangles) and T = 15 K, 0 T (green open squares) demonstrating an overall quantitative agreement of the intensity and width of the Γ point continuum. The 15 K data is scaled by a factor of 2 to account for using a different chopper frequency (see Methods). In (c) and (d) the solid lines are least-square fits to a Gaussian peak plus a constant background. **(e)** The peak height (circles) and FWHM (triangles) of the continuum as a function of field at T = 2 K (solid symbols) and T = 15 K (open symbols). The field dependence hints at a discontinuous jump close to 6 T (crossover region, shaded.) Lines are guides to the eye. **(f)** AC susceptibility (Re($\chi_{AC}$)) measured at T = 2 K with a frequency of 1 kHz shows two anomalies at $B_{C1}$ = 6.1 ± 0.5 T and $B_{C2}$ = 7.3 ± 0.3 T. In all panels **(a) – (e)** the error bars represent one standard deviation assuming Poisson counting statistics.

**Acknowledgements:**

The authors acknowledge valuable discussions with Christian Batista, Huibo Cao, Matt Stone, Feng Ye, Andrey Podelsnyak and Matthius Vojta. J. K. and R. M. particularly thank John Chalker and Dmitri Kovrizhin for collaboration on closely related work. A. B. and P. K. thank S. Chi, O. Garlea, N. Helton, J. Werner, R. Moody and M. B. Stone for assistance with the measurement on HB-3, CORELLI and HYSPEC. The work at ORNL's Spallation Neutron Source and the High Flux Isotope Reactor was supported by the United States Department of Energy (US-DOE), Office of Science - Basic Energy Sciences (BES), Scientific User Facilities Division. Part of the research was supported by the US-DOE, Office of Science - BES, Materials Sciences and Engineering Division (P.K., C.A.B. and J-Q.Y.). D.M. and P.K. acknowledge support from the Gordon and Betty Moore Foundation's EPiQS Initiative through Grant GBMF4416. The work at Dresden was in part supported by DFG grant SFB 1143 (J.K. and R.M.). J.K. is supported by the Marie Curie Programme under EC Grant agreements No.703697. The data presented in this manuscript is available from the coauthors upon request.

**Correspondence to:**

banerjeea@ornl.gov (A.B.), kelleypj@ornl.gov (P.K.), naglerse@ornl.gov (S.E.N.)

*Authors AB and PK share equal credit for first authorship.*

# Supplementary Materials:

## A. The reciprocal space guide

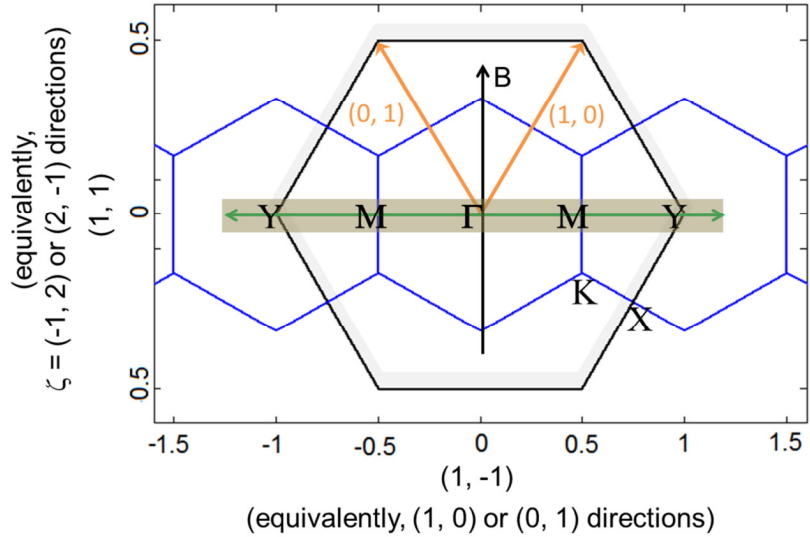

**Figure S1: The extent of the data in the reciprocal space:** The figure shows the Brillouin Zone (BZ) for the honeycomb lattice within the trigonal (*P*3 type, or *R*3 type space groups) representation following Ref. [12, S1]. In the notation used in this plot, the orange lines are the two principal reciprocal vectors (same as $a^*$ and $b^*$ in Ref. 12) denoted (1, 0) and (0, 1), respectively in 2D. The blue and the black hexagons represent the boundaries of the first and the second BZ, respectively. The black arrow marks the direction of the field ($\vec{B}$) along (1, 1). This direction is equivalently the (-1, 2), (1, -2), (-2, 1), (2, -1) or the (-1, -1) direction in the trigonal notation, with units of 2.1 Å$^{-1}$. The measurements shown in figure 2 were performed spanning Y-M-Γ-M-Y points, perpendicular to the magnetic field ($\vec{B}$), and along the direction (1, -1). The later is equivalent to the (1, 0), (-1, 0), (0, -1), (0, 1) or the (-1, 1) direction in the trigonal notation, with units of 1.21 Å$^{-1}$. The shaded region show the integration region over range Δζ = [-0.05, 0.05] marking the extent of in-plane integration range perpendicular to (H, 0) used in Figs. 2, 3 and also Fig. S4 below. In the convention chosen in this manuscript (same as Ref. 14), the reciprocal L direction has units of 0.37 Å$^{-1}$.

## B. Behavior of in-plane vs. out-of-plane magnetic Bragg peaks

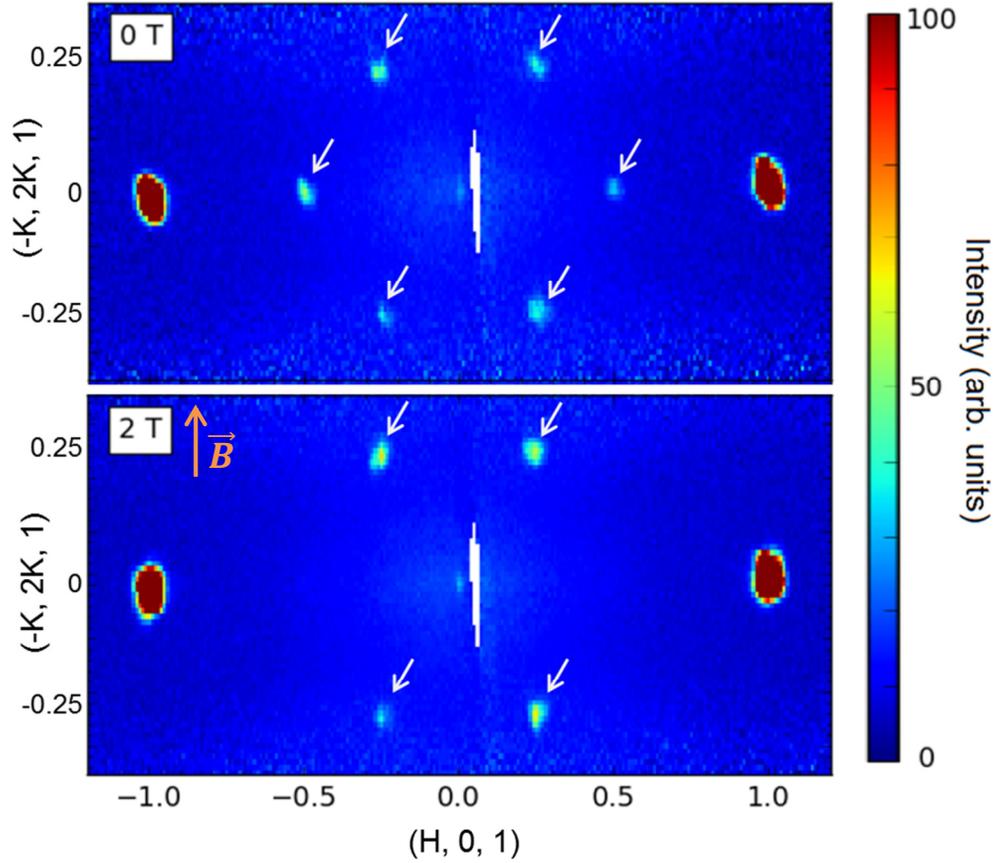

**Figure S2: In-plane vs. out-of-plane magnetic Bragg peaks at 0 and 2 Tesla.** (a) and (b) show data taken at the white-beam Laue instrument CORELLI integrated along L over the interval ΔL = [0.92, 1.08]. (Upper panel): The data taken at 0 T, 2 K showing six magnetic Bragg peaks (marked in white arrows) at the six effective M point locations (see Fig. S1 as a reference to the reciprocal lattice) (Lower panel): The data taken at 2 T (field direction marked in figure in orange arrow), 2 K showing that the in-plane peaks along (1, -1, 0) directions at the locations (½, -½, 1) and (-½, ½, 1) have completely vanished, while the magnetic Bragg peaks in the out-of-plane locations (½, 0, 1), (0, ½, 1), (0, -½, 1) and (-½, 0, 1) have enhanced. The field dependent evolution of the magnetic Bragg peaks is presented in Fig. 1c.

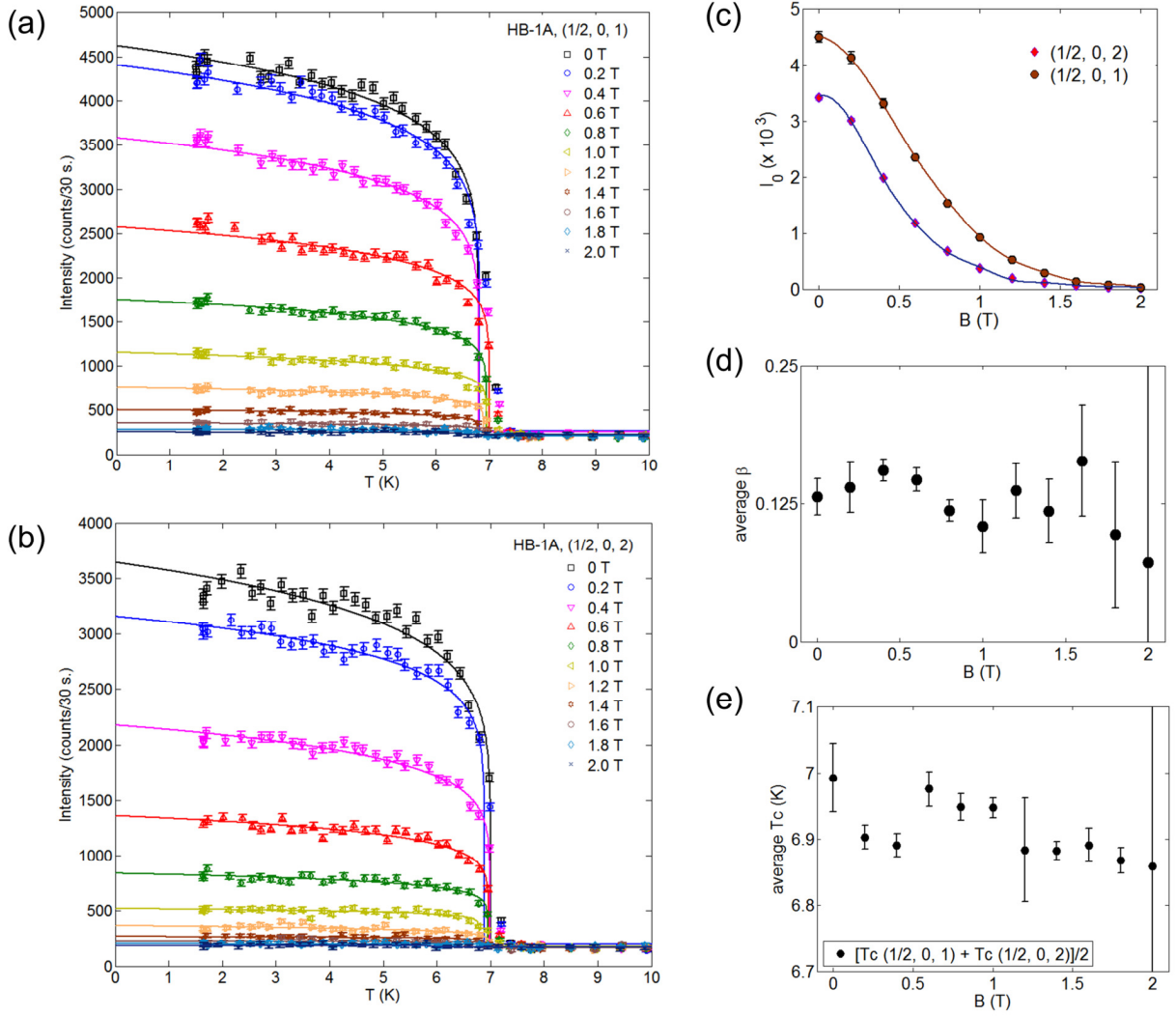

**Figure S3: Behavior of the in-plane magnetic Bragg peaks with field between 0 and 2 T.** (a) The data taken at HB-1A between 0 and 2 Tesla showing the evolution of the (a) (1/2, 0, 1) and (b) (1/2, 0, 2) magnetic Bragg peaks. The data is fit to the functional form $I = I_0 \left(1 - \frac{T}{T_c}\right)^{2\beta}$ (solid lines). The resultant $I_0$, b, and $T_c$ are plotted in panels (c), (d) & (e) respectively. The $T_c$ and the exponent $\beta$ have only a feeble field dependence in this range, with average $T_c$ = 6.91 ± 0.04 K, and average $\beta$ = 0.126 ± 0.03. These values match zero-field results reported in a different sample before [11, 12]. The lines in panel (c) are splines serving as guides to the eye. Errorbars represent 1 σ.

## C. Energy profile of spin-waves at various Q.

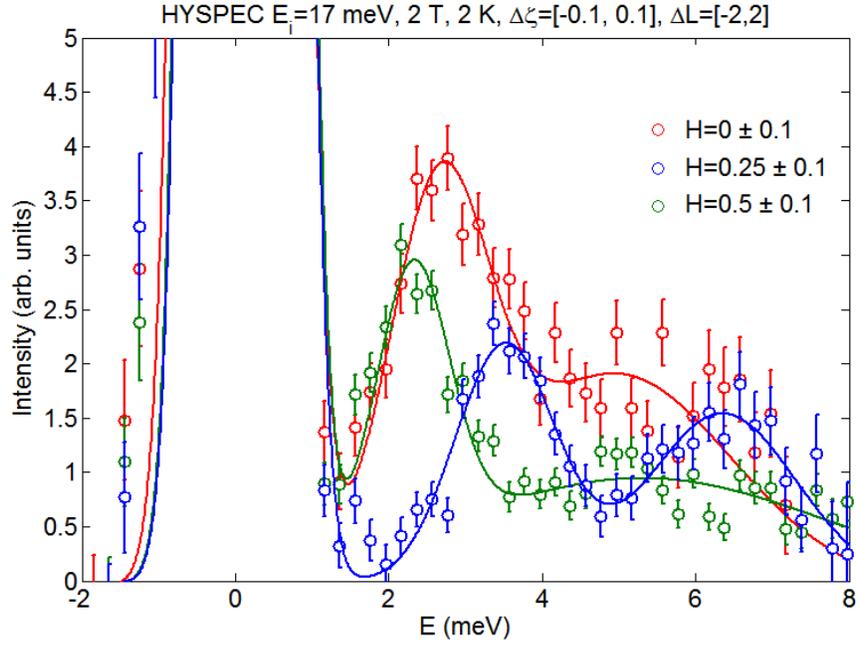

**Figure S4: Plots of various constant Q cuts on the 2 K data** obtained with $E_i$ = 17 meV at HYSPEC are shown at the Γ-point (H=0, red), M-point (H=0.5, green), and an intermediate point between the Γ and M points (H=0.25, blue) which corresponds to the antiferromagnetic zone-boundary. For best statistics, the data are shown for 2 T where the spin-wave intensity is the maximum. The data are integrated in the same region as shown in Fig. S1 in plane, with ΔL = [-2, 2], and a width of 0.2 along the (H, 0, 0) direction. The lines are Gaussian fits serving as guides to eye. The peak in the spectrum at the Γ-point occurs at 2.7 ± 0.1 meV matching THz measurements [28]. The peak in the spectrum at the M-point occurs at 2.2 ± 0.1 meV close to the value reported in Ran *et al.*, [27] while the peak in the spectrum at the intermediate point occurs at 3.5 ± 0.3 meV.

**D. Theory of the dynamical response of the Kitaev QSL.**

Here we provide details of our theoretical calculation of the dynamical response of the Kitaev QSL. The Kitaev model stands out as an example of a non-trivial interacting spin system beyond one dimension for which even dynamical properties can be calculated exactly [15-17]. The actual Hamiltonian describing the material comprises additional interactions beyond the Kitaev exchange; these are not generally agreed upon but at any rate, for a generic model Hamiltonian the calculation of physical observables is typically restricted to purely numerical approaches [S2-S5] which instead suffer, e.g., from finite size effects. Hence, we follow our previous approach [2, 12, 13] of comparing INS measurements to reliable calculations of an arguably fine-tuned pure Kitaev Hamiltonian. In the following, we sketch a phenomenological extension *beyond exact solubility* of our calculations to include nonzero temperatures and the magnetic field.

The addition of a magnetic field spoils the solubility of the pure Kitaev model, but as was already noted by Kitaev [30], time reversal symmetry can instead be broken with a soluble effective three spin interaction $h_{eff}$, which may perturbatively be interpreted as an effective magnetic field. However, establishing a quantitative equivalence is not possible, not least because this also depends on the precise form of the full Hamiltonian, for example $h_{eff} \sim h^3$ for a pure Kitaev Hamiltonian but $h_{eff} \sim h\,\Gamma^2$ for a Kitaev model with a small spin-off diagonal Gamma ($\Gamma$) term [S5]. Here, we therefore concentrate on the main qualitative features of the effect of breaking TRS and use its strength as an undetermined parameter.

We have shown previously that good quantitative results of the dynamical structure factor at the Gamma point S(q=0,ω) are obtained by a scattering problem of Majorana fermions on a local pair of $Z_2$ fluxes [15]. Here we employ an extension of this idea to finite temperature by averaging this 'adiabatic response' over a disordered flux background [S6]. Approximating fluxes as non-interacting yields a thermally excited background flux density $n_F(\Delta/T)$ in terms of the flux gap $\Delta$ in the form of the Fermi function $n_F(x) = 1/(1+e^x)$. In the notation of Ref. [16] the response is obtained from:

$$S(\mathbf{q}=0,\omega) = \left\langle \sum_n \delta\left[\omega - \Delta - E_n^F\right]|X_{n,0}|^2 \left[n_M\left(-E_n^F/T\right)\right] + \sum_n \delta\left[\omega - \Delta + E_n^F\right]|Y_{n,0}|^2 n_M\left(E_n^F/T\right) \right\rangle_{n_F(\Delta/T)}$$

Figure S5 shows the resulting dynamical structure factor at the Gamma point for a system of 800 spins sampled over 2000 appropriately sampled random flux configurations. As the effective field $h_{eff}$ is increased at low temperature (T=Δ=0.065$J_K$), the response at low frequencies is suppressed, with weight being transferred into a peak which grows as it moves to higher frequencies. Although the 6T data may be slightly affected by the remnants of weak spin wave intensity, it is very interesting to note that the calculation is in good qualitative agreement with the trend of the experimental data as the field is increased.

As an aside, we also show in Fig. S6 a comparison between FM and AFM Kitaev coupling both at low/high T (solid/dashed lines for T= Δ/7.5Δ) and zero/nonzero value of the time-reversal symmetry-breaking term (black/red lines for $h_{eff}$= 0/0.1$J_K$). For both FM and AFM models an applied field results in an upward renormalization of the main peak in the data. Notably, the low-frequency portion of the FM response is most strongly temperature dependent. While considerable uncertainties with respect to the precise microscopic Hamiltonian remain, for the data presented here, the FM choice of $J_K$, resulting in a peak in the response on the low energy side, resembles the data more closely. This would then be in keeping with the idea that the dominant interaction in α-RuCl$_3$ is a FM Kitaev-type exchange [13, 20, S3, S4], which is consistent with the observed sensitivity of the magnetization to an applied magnetic field [26].

We finally emphasize that a hallmark of the Kitaev QSL in a magnetic field is provided by the appearance of Majorana flux bound states. While it might be difficult to separate a broadened peak in a small gap from the large continuum response, it would be desirable to measure at temperatures well below the flux gap and also for a number of magnetic fields above $B_c$. Combined observations of the characteristic scaling of the gap, position and intensity of the peak [S6] would be the strong signature of the non-Abelian Kitaev QSL.

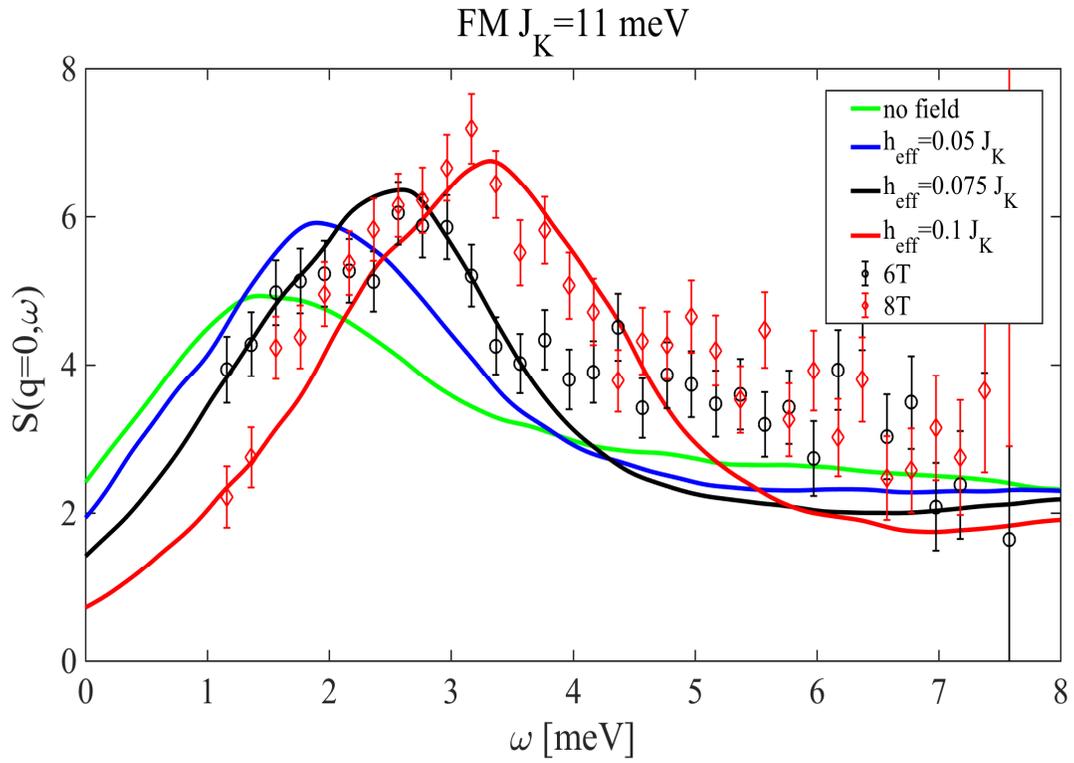

**Figure S5:** Evolution of response at Q=0 upon increasing the strength $h_{eff}$ of the time-reversal symmetry breaking term (solid lines) at a low temperature T = 0.05$J_K$. The experimental data (symbols) exhibits an analogous trend as the field is increased.

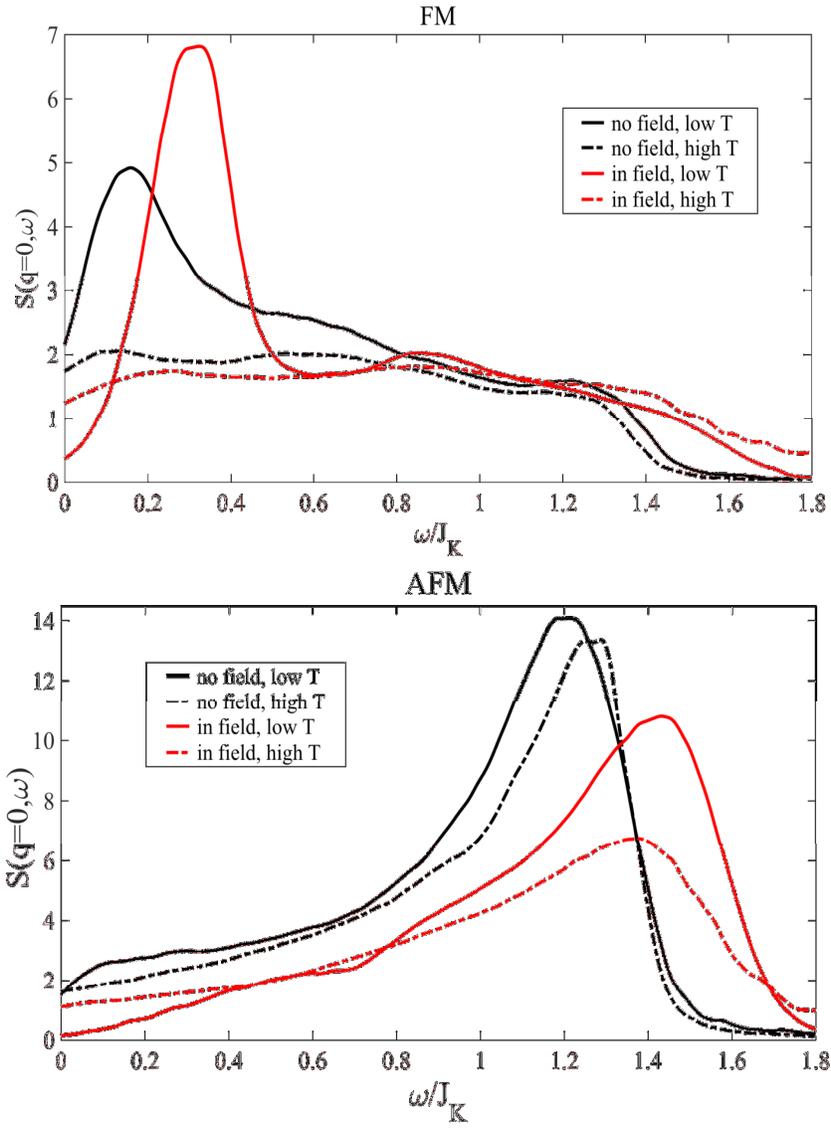

**Figure S6:** Comparison of response for FM (top panel) and AFM (bottom panel) Kitaev coupling at different temperatures and with (red) and without (black) an effective field (time-reversal breaking).

**Additional Supplementary References:**